\documentclass[12pt]{article}

\usepackage[
backend=biber,
style=numeric-comp,
sorting=none]{biblatex}
\usepackage{caption}
\usepackage{subcaption}
\usepackage{graphicx}
\graphicspath{{figures/}}

\usepackage{amssymb}
\usepackage{amsmath} 

\usepackage{algorithmic}
\usepackage{array}
\usepackage{lipsum}
\usepackage[hidelinks]{hyperref}
\begin{document}

\title{Exploring The Relationship Between Road Infrastructure and Crimes in Memphis, Tennessee} 
\author{Alexandre Signorel}
\date{December 1, 2022} 
\maketitle 

\begin{abstract}
Memphis, Tennessee is one of the cities with highest crime rate in the United States. 
In this work, we explore the relationship between road infrastructure, especially potholes, and crimes. 
The pothole and crime data are collected from~\cite{memphisdata} between 2020 and 2022.
The crime data report various crimes in the Memphis area, which contain the location, time, and type of the crime.
The pothole data is part of the Open 311 data~\cite{memphisdata}, which contains information of different infrastructure projects, including the location of the project, and the starting and ending dates of the project. We focus on infrastructure projects regarding pothole repairs.

\end{abstract}

\section{Introduction}
Memphis, Tennessee is one of the cities with highest crime rate in the United States.
In this task, we are interested in studying the potential relationship between the road infrastructure, potholes in particular, and crimes in Memphis.
The data is collected from the Memphis Data Hub~\cite{memphisdata}, 
There are three main data categories that are currently accessible: \emph{Public Safety} which contains crime data; \emph{Open 311} which contains data about current and past infrastructure projects; and \emph{Civic Assets} which contains data about public services. 

For our project, we focus mainly on Public Safety and Open 311 where in Open 311 we specifically focus on data regarding potholes. Our goal is to explore potential correlations between crimes that happened in an area and the road condition of that area, evaluated by the amount of potholes reported.

\section{Working with Data} \label{download}
The data are downloaded in the CSV format.
We process the data and extract the data from 2020 to 2022. This is the time period when both pothole data and crime data are available.

\subsection{Crime Data}
\begin{figure}[hbt!]
    \centering\includegraphics[width=\linewidth]{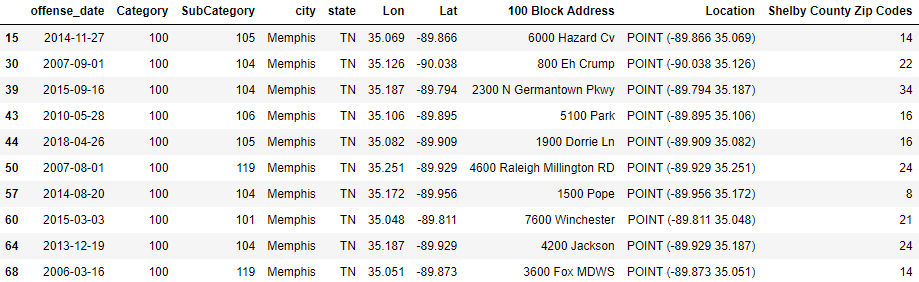}
    \caption{An example of the processed crime data.}
    \label{fig:CrimeData}
\end{figure}

To process the crime data, we convert crime names into numerical values that range from 100 to 1700. This indicates the overall 17 types of crimes. The last two digits of the numerical values indicate the sub-categories of crimes. 
Next, we remove information that does not concern the crime's location and time.
Lastly, we convert the date to the Date-Time format, which displays the year, month, and day of a crime. 
An example of the processed crime data is shown in Figure~\ref{fig:CrimeData}. 

\subsection{Pothole Data}
We extract the pothole data from the Open 311 data.
We convert the extracted date to the Date-Time format as of the pothole data.
\begin{figure}[hbt!]
    \centering\includegraphics[width=\linewidth]{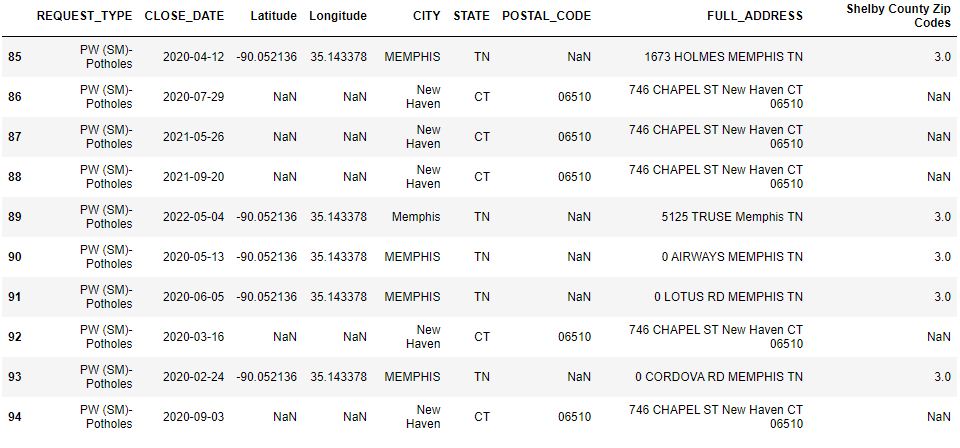}
    \caption{An example of the processed pothole data.}
    \label{fig:PotholeData}
\end{figure}

\subsection{Visualizing Data}
\begin{figure}[hbt!]
    \centering\includegraphics[width=\linewidth]{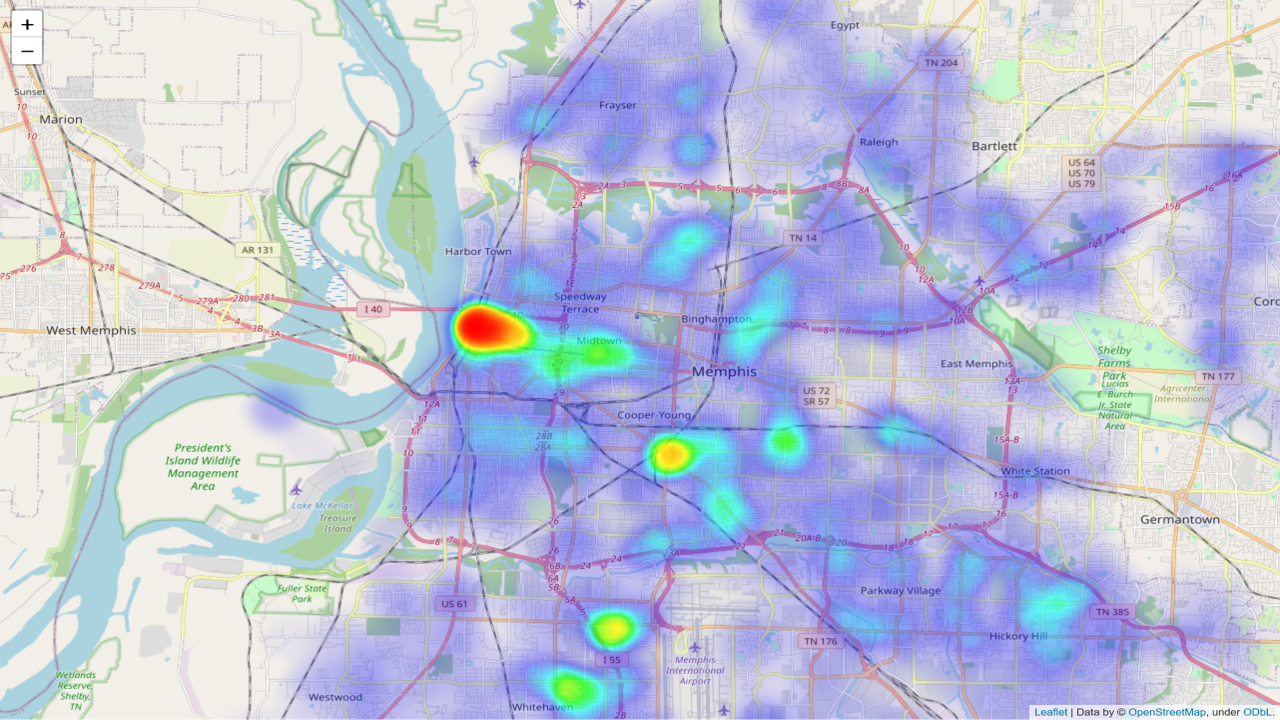}
    \caption{The heat map of crimes in Memphis between April 2020 and June 2020. }
    \label{fig:CrimeHeatMap}

    \centering\includegraphics[width=\linewidth]{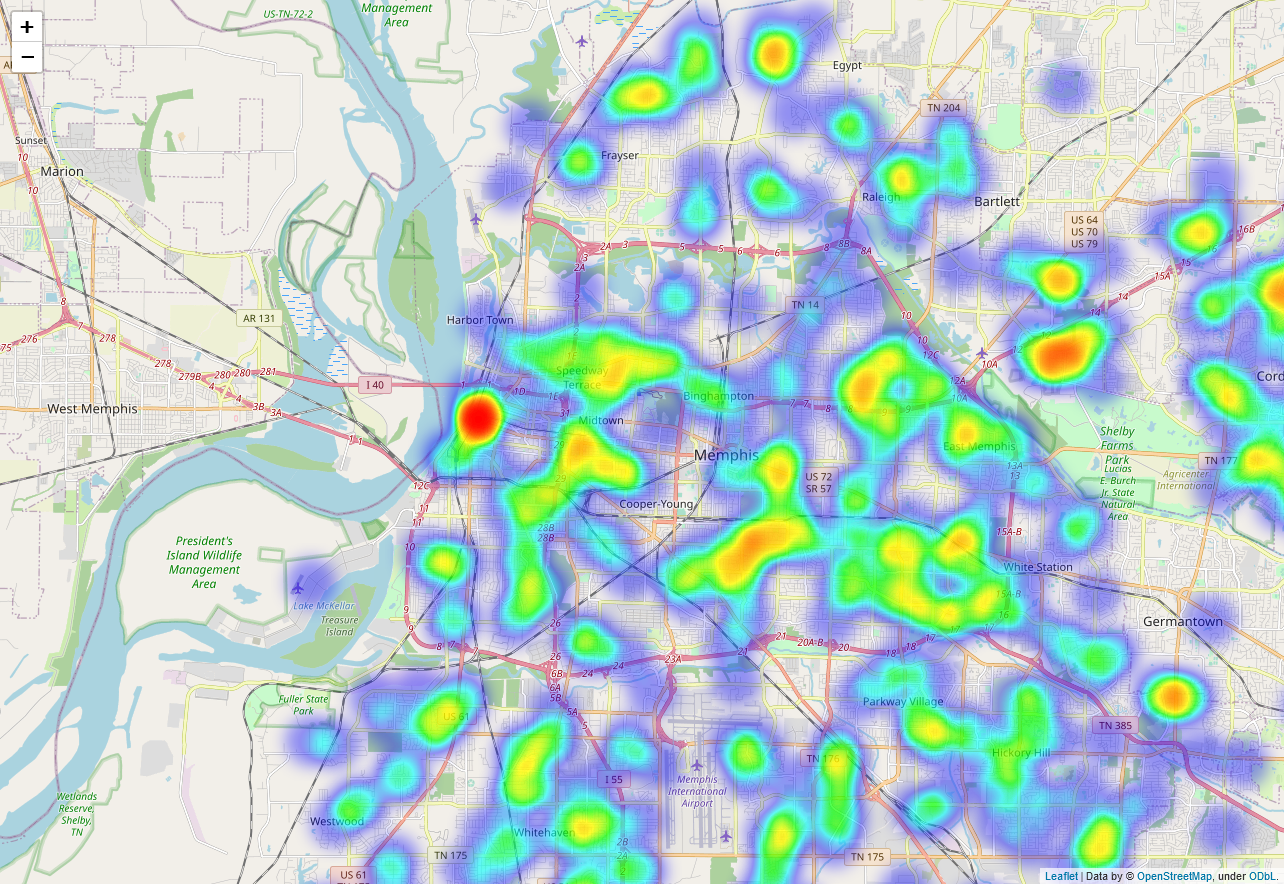}
    \caption{The heat map of potholes reported in Memphis between April 2020 and June 2020.}
    \label{fig:PotHoleHeatMap}
\end{figure}

In order to qualitatively study the data, we visualize both types of data by dividing them into windows of three months.
We then use Folium~\cite{folium}, a python package that uses OpenStreetMap~\cite{OpenStreetMap}, to generate a view of the Memphis area.
Next, we use Folium.plugins to generate a heat map which is overlaid onto the map of Memphis. 
The heat map of crimes is shown in Figure~\ref{fig:CrimeHeatMap} and
the heat map of potholes is shown in Figure~\ref{fig:PotHoleHeatMap}. 

\subsection{Analyzing Data}
\begin{figure}[hbt!]
    \centering\includegraphics[width=\linewidth]{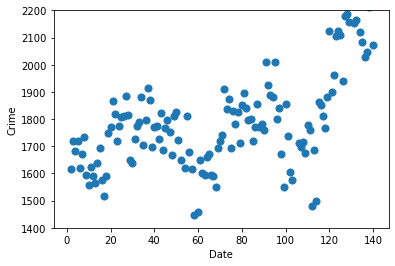}
    \caption{The scatter plot of crimes in Memphis between January 2020 and October 2022.}
    \label{fig:CrimeGraph}

    \centering\includegraphics[width=\linewidth]{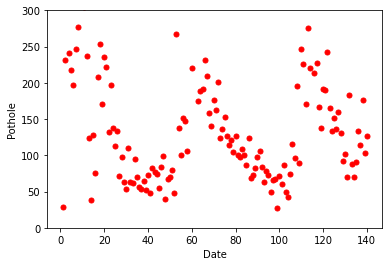}
    \caption{The scatter plot of potholes between January 2020 and October 2022.}
    \label{fig:PotHoleGraph}
\end{figure}

In order to quantitatively study the data, we convert both types of data into windows of a week between January 2020 and October 2022. 
The scatter plots of crimes and potholes are shown in Figure~\ref{fig:CrimeGraph} and Figure~\ref{fig:PotHoleGraph}, respectively. 

\section{Results} 
\label{results}
After examining the data, we find that both types of data exhibit wave patterns. 
The crime data first go up and then back down with an upward trend within the last wave.
The pothole data exhibit the wave pattern more explicitly with the lowest being around the week 40 of each year. 
In all three years, we can see that the number of crimes reaches a peak while the amount of potholes reaches a bottom. However, with the limited data and fluctuating patterns, it is difficult to claim any correlation between the two. 

\section{Conclusion and Future Work} 
In this task, we explore the relationship between potholes and crimes in Memphis, Tennessee.
We find that both types of data exhibit wave patterns. 
However, it is difficult to claim that they are strictly correlated.

In the future, we would like to expand our study to cover more infrastructure such as  buildings and public services within an area.
In addition, instead of conducting a correlation study of the entire Memphis area, we can compare higher-crime-rate areas with lower-crime-rate areas to explore the relationships between various infrastructure factors and the crimes.


\printbibliography

\end{document}